# Tuning residual stress, directional memory and aging in soft glassy materials

Paolo Edera[(1)], Minaspi Bantawa[(2)], Stefano Aime[(1)], Roger T. Bonnecaze[(2)], Michel Cloitre[(1)]

(1) Molecular, Macromolecular Chemistry and Materials, ESPCI Paris, CNRS, PSL Research University, 75005 Paris, France

(2) McKetta Department of Chemical Engineering and Texas Materials Institute, University of Texas at Austin, Austin, TX 78712, USA

**Abstract**

When glassy materials are rapidly quenched from the liquid to the solid state upon flow cessation or cooling, they solidify in an out-of-equilibrium configuration, retaining the memory of the processing conditions for very long times. This is the origin of various phenomena, such as residual stresses and directional memory, which greatly affect their properties. At the same time, annealing the mechanical history encoded in disordered materials constitute a great challenge. Here, we address this problem for the case of colloidal glasses made of soft particles densely packed at a high volume fraction, using experiments and particle dynamic simulations. We demonstrate that periodically training soft particle glasses with a sequence of stress-controlled oscillations successfully anneals residual stress and directional memory when the stress amplitude corresponds to the yield point. At the microscopic level, annealing provides a fine tuning of the local distribution of the stress carried by the particles. Through the simulations, we show that the first moments of this distribution have precise physical meaning: the mean value of the distribution corresponds to the macroscopic stress; the skewness carries information about directional memory; and the standard deviation is related to mechanical aging. The same methodology is successfully applied to silica gels with thixotropic properties, suggesting that it is general and may be extended to other classes of disordered materials.

**Relevance**

Glassy materials have the capacity to remember the mechanical history they experience during their processing. They can trap long-lived residual stresses, store directional memories, and undergo aging. These different phenomena are not well understood, they are generally entangled and they strongly impact the final properties. Controlling and erasing memory effects remains still a great challenge. Here we address this problem in the case of concentrated colloidal pastes with the support of experiments and particle dynamic simulations. We demonstrate that cyclic shear deformations with well-chosen amplitudes enable the successful preparation of pastes without residual stresses or directional memory, and with tunable age. Annealing is associated, at the microscopic scale, with a fine tuning of the distribution of local stresses.

# Introduction

Many disordered materials of practical interest are processed in the liquid state and rapidly quenched to the solid state upon flow cessation or cooling. Upon solidification, the microstructure is instantaneously frozen in an out-of-equilibrium configuration and mechanical stresses remain trapped in the materials. The existence of residual stresses has been described and analyzed in systems as different as metallic glasses (1), Gorilla® glass (2), hard sphere suspensions (3), jammed suspensions (4), and biological materials (5). Over time, the metastable microstructure undergoes slow rearrangements through a process known as aging that progressively lowers the free energy, increases the stiffness, and eventually relaxes the residual stress, at least partially. Since aging involves the existence of very long relaxation processes associated with frustration and cooperativeness, most materials do not reach mechanical equilibrium over accessible times scales, resulting in persistent residual stresses.

Residual stresses and aging are often associated with memories. Shearing materials along a direction can encode a directional memory: the material remembers its previous mechanical history. An important consequence is that the response to an external drive is not symmetric upon flow reversal. For instance, a material initially sheared along one given direction appears stiffer in this direction than in the opposite direction because of the action of accumulated restoring stresses. Directional memory has been described in many disordered materials including jammed suspensions (6,7,8), metallic glasses (9), granular materials (10,11), nanocomposites (12), and colloidal gels (13). It is of outstanding importance in many respects. One manifestation, known as the Bauschinger effect in the field of crystal and metal plasticity, leads to a dramatic increase of the elastic modulus and of the yield stress (14). Directional memory is also at the origin of the spontaneous motion occurring without forced flow in colloidal glasses and biological materials (5). Finally, it adversely affects the physical characterization of materials (13) and can be a major issue in processing design and operation (15).

Recently, the topic of memory formation in disordered materials has attracted a lot of attention, and many forms of memories have been identified (16). Besides directional memory induced by a unidirectional deformation, repeated cycles of deformation can also encode a memory of the amplitude of deformation in athermal disordered materials. Cyclically training a material using an oscillatory shear deformation writes a memory of the shearing amplitude, which can be subsequently read by a sequence of oscillations with increasing amplitude (17,18). The possibility to encode multiple memories in a system and the comparative merit of various write-and-read protocols have received increasing attention over the last years (19,20,21). However, to date, the possibility to prepare materials free of residual stresses and memory effects is highly desirable and remains a great challenge (13).

Here, we address this question using Soft Particle Glasses (SPG), which can be considered as an archetype of athermal glassy materials. SPGs are dense assemblies of soft and deformable particles packed in an amorphous state at high volume fraction. An experimental realization of SPG is a suspension of polyelectrolyte microgels swollen in water at concentrations above the jamming transition (22). The particles have a well-defined microstructure and interact through purely repulsive contact forces of elastic origin. First, we show that training SPGs periodically with a sequence of well-chosen stress-controlled oscillations erases any directional memory originating from the mechanical history of the materials. The optimum shearing conditions are obtained when the applied stress is close to the yield stress. Secondly, we use particle dynamic simulations on model SPGs to analyze the microscopic origin of directional memory in these materials. Simulations quantitatively reproduce the macroscopic behavior found in experiments and give access to the particle stress distribution, which we express in terms of the probability density function $P(\sigma_p)$ where $\sigma_p$ is the local contribution of the stress associated to a single particle, and its evolution over time. We demonstrate that the directional memory originates from the asymmetry of $P(\sigma_p)$, parametrized by the skewness parameter, which progressively decreases during training as the particles contributing to the tails of the distribution yield. The width of $P(\sigma_p)$, parametrized by the standard deviation, becomes narrower as aging takes place. We are thus able to disentangle directional memory and aging. We show that this protocol to anneal the macroscopic residual stress and the directional memory applies to other types of amorphous materials, like colloidal gels made of fume silica in an organic solvent (13).

**Directional memory in soft particle glasses**

Preshearing prior to rheological measurements is a common strategy to prepare soft glassy materials in a reproducible state, irrespective of their previous mechanical history. While this procedure is believed to rejuvenate the material by breaking the structure formed over time, it encodes a mechanical memory that manifests either as a time-dependent strain recovery at zero imposed stress, or as stress growth at constant imposed deformation, as shown in Figs 1A-B. In this experiment, a microgel suspension in the jamming regime (see Materials and Methods) is presheared at $\dot{\gamma} = 20 \ s^{-1}$ for a time interval of 50 s. At $t = 0$, the preshear flow is stopped, the strain origin being reset to zero, and the applied stress is set to zero. We observe that the material is slowly pulled back on itself in the reverse direction; this indicates that is has developed a directional memory during the preshear. The effect is quantified in Fig.1A, where we plot as blue points the time evolution of the strain during the recovery: the accumulated strain slowly decreases but does not reach a constant value over a period of time as long as $10^4$ s as depicted by the thin line in Fig. 1A). Regardless of the preshear direction, the recovery is always in the opposite direction.

Since the total recoverable strain cannot be reached in this experiment, we interrupt the recovery at $t_1 = 80$ s, imposing that the mobile geometry of the rheometer remains motionless for $t > t_1$ ($\dot{\gamma} = 0$ $s^{-1}$), as indicated by the black data points in Fig. 1A. The stress, which was set to zero during the recovery, starts increasing approximately as a logarithmic function of time, as shown by the black data points in Fig. 1B. This experiment proves that setting the macroscopic stress to zero is not sufficient to erase directional memory. While in the equilibrated state, the net forces and torques acting on each particle are negligible, this memory effect implies that the preshear flow creates long-lived local structural distortions that are responsible for an imbalance of the local strains (when $\sigma = 0$) or stresses (when $\dot{\gamma} = 0$).

The long-time directional memory encoded during preshear dramatically impacts subsequent measurements. This is demonstrated by performing stress-growth experiments, consisting in applying a constant shear rate of magnitude $|\dot{\gamma}_0| = 1$ $s^{-1}$ at $t = t_1$, either in the direction of preshear (hereafter named forward direction) or in the opposite direction (backward direction). We find that in both cases, the absolute value of the stress increases from zero, passes through a maximum called the static yield stress, and slowly decreases to its steady state values, as shown in Fig. 1C. The static yield stress represents the stress barrier that the material overcomes before reaching steady flow. Remarkably, the stress evolutions for the forward and backward directions do not superimpose. In particular, we find that the static yield stress is systematically larger for the forward direction. This observation recalls the Bauschinger effect that takes place during the loading and unloading of crystalline and glassy metals (14). To quantify the asymmetry between the forward and backward stress-growth experiments we define the asymmetry parameter $A$ as the area between the two curves over a fixed interval encompassing the entire transient responses:

$$A = \int_0^2 \left[ \sigma(\dot{\gamma}_0) - |\sigma(-\dot{\gamma}_0)| \right] d(\log_{10}(\gamma)) \qquad (1)$$

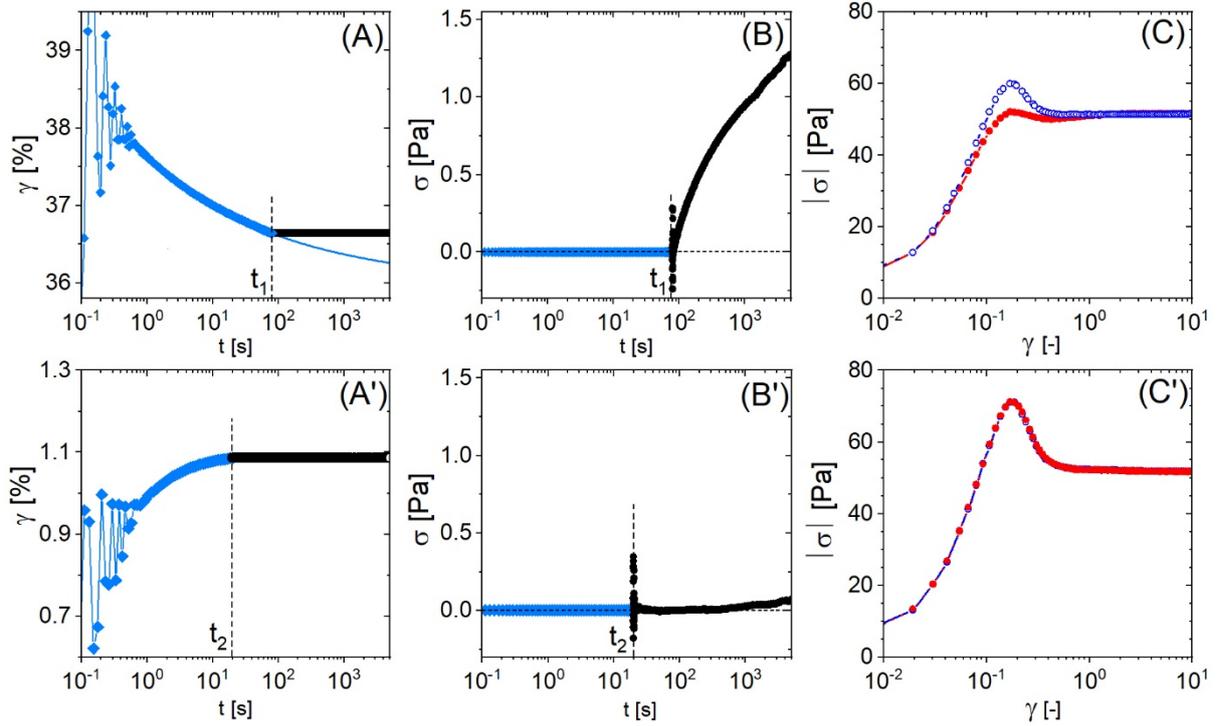

**Fig. 1.** Strain recovery (A, A'), stress relaxation (B, B'), and stress growth experiments (C, C') after preshear flow (top) and optimal oscillatory annealing (bottom). Preshear is a continuous shear flow ($\dot{\gamma} = 20\ s^{-1}$) of duration 50 s (not shown). Annealing uses two sequences of stress-controlled oscillations ($\omega = 1$ Hz) applied for 30 s each, at stress amplitudes $\sigma = 100$ Pa (conditioning) and $\sigma = 31$ Pa (training). In (A) and (B), the stress is set to zero at the time origin taken at the end of preshear; the strain is maintained constant after $t_1 = 80\ s$. In (A') and (B'), the time origin is at the end of annealing; the stress is first set to 0 and at $t_2 = 20\ s$, the strain is maintained constant. Note that $t_1 = 80$ s is the sum of $t_2 = 20$ s and the duration of the annealing interval (60 s). The stress growth experiments are performed by applying two opposite shear rates $\dot{\gamma}_0 = +1\ s^{-1}$ (blue curve) and $\dot{\gamma}_0 = -1\ s^{-1}$ (red curve) at $t_1$ and $t_2$, respectively. The material is a microgel glass with a low-frequency plateau modulus $G_0 = 435$ Pa, a yield stress $\sigma_y = 31$ Pa, and yield strain $\gamma_y = 31$ Pa (Materials and Methods and *SI Appendix*)

$A$ decreases very slowly with increasing waiting time $t_1$, yet it does not vanish for $t_1$ up to $10^4$ s, in line with previous reports (6). This motivates the search for efficient ways to relax macroscopic residual stresses and erase the directional memory imprinted by the preshear flow.

## Annealing using oscillatory training

Repeated shear oscillatory deformations have the capacity to imprint persistent information in athermal amorphous materials. Simple and multiple memories of amplitude shear deformations have been encoded and subsequently read by successive cyclic shearing operations. This prompts the reverse question of whether it is possible to prepare materials without macroscopic residual stress and directional memory using periodic shear deformations. To address this question, we first perform oscillatory strain sweep measurements, keeping the frequency constant ($\omega = 1$ Hz) and increasing the stress amplitude. The observed behavior is characteristic of that found for many soft glassy materials (22). The stress amplitude, $\sigma_0$, first increases linearly with the strain amplitude, $\gamma_0$, signaling the linear response regime; linearity breaks down at some point above which $\sigma_0$ varies as a power law of $\gamma_0$ with a weak exponent (*SI Appendix,* Fig. S1). This behavior denotes the flow regime. The intersection between the linear response regime and the flow regime is well defined and corresponds to the yield point. Figure 2A shows the strain-stress when the stress amplitude is scaled by the yield stress ($\sigma_y = 31$ Pa) and the strain amplitude by the yield strain ($\gamma_y = 0.075$). It has been shown independently that the yield point in soft glasses is associated with a dynamical transition at the microscopic scale: at small strain or stress amplitudes, the displacements of the particles are localized, whereas above the yield point, shear irreversible displacements leading to shear-induced diffusion take place (23-26). Moreover, the vicinity of the yield point is singular since it is where cooperative relaxation leading to large scale spatial rearrangements is maximal (27, 28). We thus hypothesize that periodic oscillations applied close to the yield point can reorganize the microstructure in such a way that the asymmetry introduced by the directional memory is erased.

To test this hypothesis, we have designed the experimental protocol schematized in the inset of Fig. 2A. First, the microgel glass is presheared at $\dot{\gamma} = 20 s^{-1}$ as above, to imprint a mechanical history. Our new protocol comprises three successive steps, called conditioning, training, and reading. Conditioning and training involve series of 30 stress-controlled oscillations at constant frequency $\omega = 1$ Hz. The choice of controlling the stress instead of the strain is motivated by the lack of well-defined reference strain: indeed, the recoverable strain is not known a priori and the strain is endlessly changing with time. As a conditioning step, we choose a sequence of large amplitude oscillations ($\sigma_c = 100$ Pa), which fully fluidizes the material and puts it in a reproducible state irrespective of the previous shear history. The subsequent training step is a sequence of oscillations at smaller stress amplitude ($\sigma_t = 31$ Pa). Its role is to reorganize the microstructure in order to erase directional memory and restore symmetry upon flow reversal. Finally, the reading step is intended to assess the performance of the training step. It is achieved by performing two stress-growth experiments in the forward ($+\dot{\gamma}_0$) and backward directions ($-\dot{\gamma}_0$), from which the asymmetry parameter *A* is computed ($\dot{\gamma}_0 = 1$ s$^{-1}$).

Fig. 2B shows that the asymmetry parameter varies with the stress amplitude of the training sequence in a non-monotonic way. For convenience and for the sake of comparison with Fig. 2A, the data are plotted as a function of the reduced strain, $\gamma_0/\gamma_y \gg 1$, even though the protocol is stress-controlled. In line with our working hypothesis, we observe a deep and fairly broad minimum for amplitudes of the training sequence close to the yield value, confirming that oscillatory shearing contributes to erasing directional memory. Moreover, this contribution is maximal around the yield point. In the flow regime $\gamma_0/\gamma_y \gg 1$, oscillations induce irreversible rearrangements that lead to large asymmetry parameters, revealing directional memory. Conversely in the linear regime ($\gamma_0/\gamma_y \ll 1$), the asymmetry persists because the amplitude of the oscillations is too small to erase the memory of preshear flow and of conditioning.

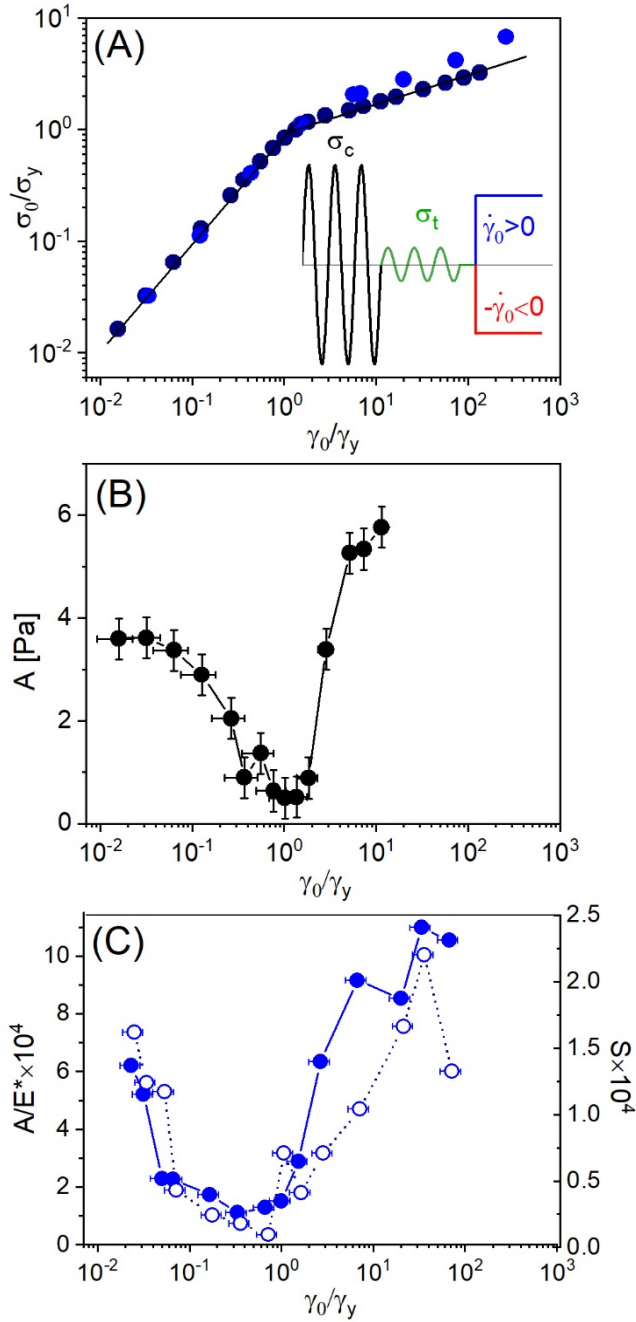

**Fig. 2**. (A) Large Amplitude Shear Rheology of a microgel glass (experiments; ●) and a model SPG (simulations; ●). The stress amplitude $\sigma_0$ and the yield strain amplitude $\gamma_0$ are scaled by the yield stress $\sigma_y$ and the yield strain $\gamma_y$, respectively. The inset depicts the controlled-stress annealing protocol used in this study. (B) Asymmetry parameters measured for a jammed microgel suspension (experiments). (C) Asymmetry parameters (●) for the model SPG (simulations) and skewness parameters (O) of the local stress distributions.

To further evaluate the quality of our annealing protocol, we perform the same experiments as in Fig. 1A-B after annealing using a training stress equal to the yield stress. This time, under zero imposed stress, the strain quickly reaches a small constant value of about 1% as shown in Fig. 1A'. Furthermore, as shown in Fig. 1B', the stress remains close to zero when the strain is maintained constant and equal to its plateau value, showing that there is no residual stress trapped inside the glass. In Fig. 1C', the stress-growth functions measured in the forward and backward directions are exactly superimposed. We have checked that the parameter $A$ is of the order of experimental uncertainty estimated from two identical experiments ($A \cong 0.4$ Pa) This result, obtained with a protocol that lasts about one minute, exceeds what could be achieved by keeping the materials at rest for hours. To address the generality of our results, we have reproduced the experiments used to build Fig. 2B using strain-controlled oscillatory deformations and a simplified annealing protocol that omits the conditioning step. The results, reported in *SI appendix,* Fig. S2 demonstrate that the asymmetry parameter has a well-pronounced minimum near the yield point in all cases. However, the minimum is deeper and narrower when conditioning and training are combined and stress-controlled oscillations are applied. A complete understanding of these results is beyond the scope of this contribution and will be reported elsewhere.

**Microscopic origin of residual stress and directional memory**

To elucidate how jammed suspensions of soft particles remember and forget their past history, we now perform particle dynamic simulations involving a model SPG made of non-Brownian elastic spheres dispersed in a Newtonian fluid above the jamming transition (Materials and Methods) (Fig. 3A). The particles interact elastically through Hertzian contacts characterized by the contact modulus $E^* = E/(1 - \nu^2)$, where $E$ is the Young modulus and $\nu$ the Poisson ratio (29). It is important to note that the interaction between the microgels in the jammed state is well described by this Hertz potential; the contact modulus for microgels has been estimated to be of the order of $E^* = 10^4$. We refer the reader to the Materials and Methods section for a brief description of the simulation technique.

First, we investigate the yielding behavior of the model SPG under oscillatory shear deformation using strain sweep simulations. The strain-stress curve has the same characteristic form as in experiments, allowing us to determine the non-dimensional yield stress $\sigma_y/E^* = 0.0018$ and yield strain $\gamma_y = 0.031$ (*SI Appendix,* Fig. S1 and Table S3). Fig. 2A shows that the non-dimensional strain-stress curves computed in simulations and measured experimentally are in good agreement. We capitalize on this result to perform simulations that mimic the experiments reported previously. To imprint directional memory, the SPG is subjected to a preshear flow that is characterized by the non-dimensional shear rate $\dot{\gamma}\eta_S/E^* = 10^{-6}$. We use the simplified annealing protocol, training the SPG with a sequence of 20 oscillations at constant frequency $\omega\eta_S/E^* = 10^{-6}$, which are applied in strain-controlled mode. As before, reading consists of two stress-growth experiments in the forward and backward directions ($|\dot{\gamma}\eta_S/E^*| = 10^{-6}$), from which the asymmetry parameter is computed. The results are shown in Fig. 2C where the dimensionless asymmetry $A/E^*$ is plotted as a function of the non-dimensional strain amplitude $\gamma_0/\gamma_y$. Just as found in experiments, the asymmetry parameter exhibits a clear minimum in the vicinity of the yield point, indicating that the directional memory has been minimized. This is confirmed in Fig. 3B, where the stress-growth functions in the forward and backward directions are well-superimposed when the strain amplitude of the training sequence is equal to the yield strain.

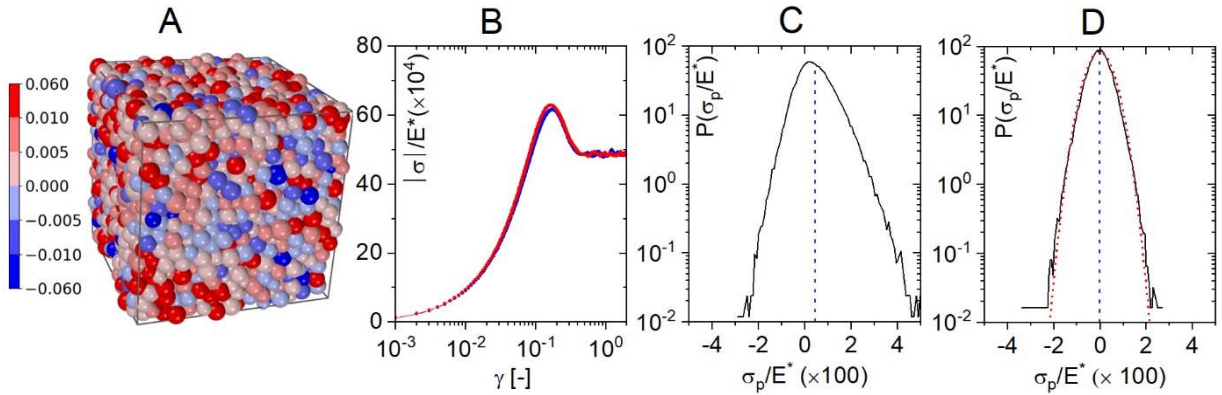

**Fig. 3** (A) 3D visualization of the simulation box showing local shear stress values on individual particles just after flow cessation ($\dot{\gamma}\eta_S/E^* = 10^{-6}$), with particles colored in red (blue) bear positive (negative) stress. (B) Simulated stress growth experiments in the forward (blue) and backward directions (red) after oscillatory training at $\gamma_0 = \gamma_y$ and $\omega\eta_S/E^* = 10^{-6}$. (C, D) Probability density functions (PDF) of particle shear stresses immediately after preshear flow and after optimum annealing, respectively. The blue dotted lines indicate the mean values of the PDF; the red dotted line in (D) is the Gaussian function that best represents the PDF.

We are now able to analyze the existence of local stress and strain distributions, which was suggested by the experimental results reported in Figs 1A-B. Fig. 3A shows that the particles just after the preshear flow carry local shear stresses ($\sigma_p$) which can be positive or negative and are widely distributed. This observation is quantitatively supported by the probability density function $P(\sigma_p/E^*)$ shown in Fig. 3C. The function, which is not Gaussian, can be characterized by its statistical properties. The mean value represents the macroscopic residual stress after flow cessation. It is interesting to note that the local stresses can take values which are far different from the mean value. Moreover, there is an excess of particles carrying a positive stress leading to a positive asymmetry that can be characterized by the skewness of the distribution, $S$ (Materials and Methods). In Fig. 2C, we track the value of the skewness at the end of the annealing protocol as a function of the strain amplitude during the oscillatory training. The variations of the skewness $S$ and the asymmetry parameter $A/E^*$ can be mapped one on the other, demonstrating that $S$ is a powerful indicator of directional memory at the microscopic level. After successful annealing, the probability density function acquires a symmetric Gaussian shape centered around zero, expressing the absence of residual stress and directional memory.

It is interesting to note that, upon annealing, the particle stress distribution becomes narrower with a smaller standard deviation, suggesting that training progressively reduces the stress fluctuations inside the SPG. We investigate this effect by training the glass with an increasing number of cycles keeping constant the frequency ($\omega\eta_s/E^* = 10^{-6}$) and the strain amplitude ($\gamma_0 = \gamma_y$). The results are shown in Fig. 4A. The standard deviation, plotted in the inset, decreases in a logarithmic way with the number of cycles, indicating that the stress distribution keeps on changing very slowly after directional memory has been erased. The logarithmic form of the decay suggests that the slow evolution that is found correspond to an aging process taking place under deformation (30). In the main graph, we plot the variations of the transient stress computed during reading. The stress overshoot increases steadily with the number of cycles applied during training. To interpret this result, we revisit the origin of the stress overshoot in stress-growth experiments: it is associated to a transient accumulation of elastic contacts in the compressive region of the deformation field due to the limited mobility of the particles in the early stage of flow inception (6, 31). In other words, higher stress overshoots mean greater accumulation of contacts and slower particle mobility. We thus conclude that the relaxation of stress fluctuations during annealing decreases the local mobility of the particles and slow down rearrangements. In Fig. 4B, this interpretation is corroborated by the results of stress-growth experiments performed on the microgel glass after optimum annealing using an increasing number of cycles in the training step, $N_C$. The experimental results are in agreement with the simulations (with $E^* = 10^4$ Pa) and again we observe that the stress overshoot increases with $N_C$.

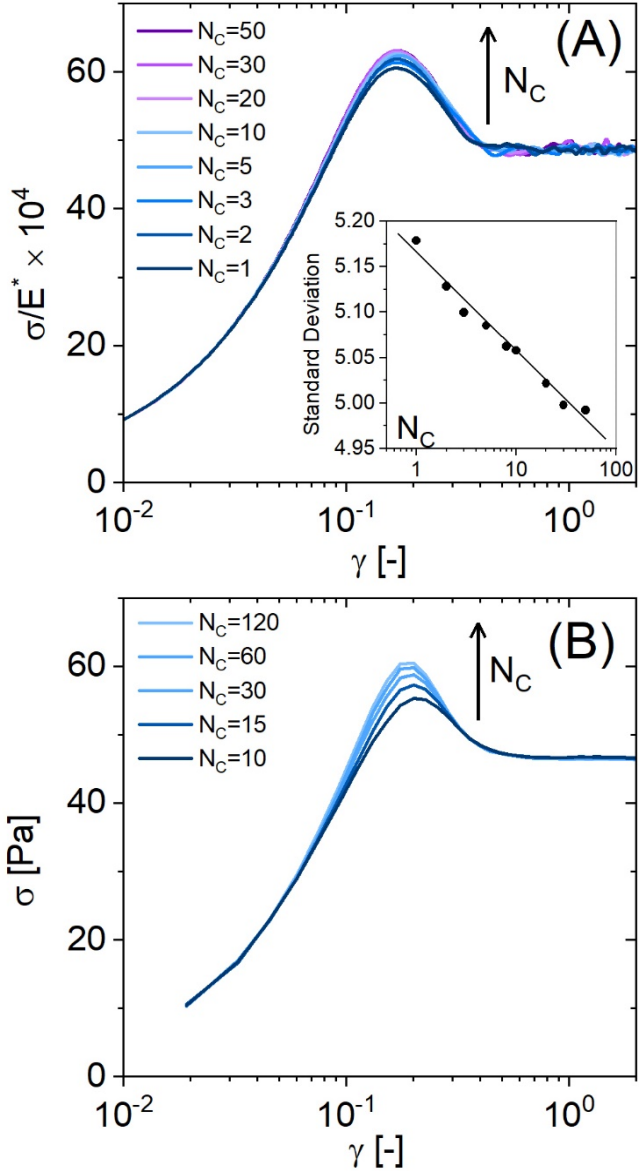

**Fig. 5** (A) Stress-growth simulations for the model SPG as the numbers of cycles applied during training, $N_C$, increases ($\omega\eta_S/E^* = 10^{-6}$ and $\gamma_0 = \gamma_y$); the inset shows the decrease in the standard deviation of the probability density function with increasing $N_C$. (B) Stress-growth experiments for the microgel glass with increasing $N_C$.

## Discussion

Our results demonstrate that the existence of residual stress, directional memory, and aging in athermal soft glasses takes its origin in the properties of the stresses imprinted at the particle scale during the mechanical history of the material. Subjecting a soft glass to a macroscopic shear deformation results in a distribution of local stresses (32), which is non-zero in average, broad due to large stress fluctuations, and asymmetric with an excess of particles with positive or negative stresses according to the flow direction. When the shear deformation is stopped, the microscopic stress distribution remains trapped because the relaxation of particle configurations involves cooperativeness, frustration, and metastability (4, 33). The memory of the material after flow cessation is entirely embedded in the properties of the local stress distributions: the mean particle stress represents the residual stress stored inside the material, the asymmetry that can be quantified by the skewness characterizes directional memory, the standard deviation is associated with aging. This conceptual framework complements previous interpretations that rely on the analysis of distortions of the average microstructure (4,6,34).

Preparing amorphous materials free of directional memory and residual stress is a true challenge because memory effects are of elastic origin, and annealing involves some degree of plasticity to relax trapped stresses. We have demonstrated that oscillatory training applied near the yield point constitutes an efficient way to erase directional memory. Our interpretation relies on the experimental observation that the vicinity of the yield point marks the transition between a reversible quiescent state and irreversible shear-induced motion and is associated to significant cooperative relaxation (18). When a material with an asymmetric local stress distribution is exposed to cyclic shear deformation near this point, only particle configurations contributing to the tails of the stress distribution are driven beyond the local yield stress and partially relax their stress. If training is prolonged after directional memory has been erased, the tails of the distribution are slowly eroded in a symmetric way, which decreases the elastic energy stored in the material and slows down further rearrangements. Remarkably, these properties can be quantified by statistical indicators of local properties without any obvious signature in global quantities like the mean-square displacements (19).

An interesting question concerns the location of the optimum shearing point. In the literature, the reversibility-irreversibility transition is often associated with the yield point (17,18,35). In our experiments, optimum shearing is signaled by a minimum of the asymmetry parameter $A/E^*$ around the yield point. However, the shape of the minimum and the quality of annealing depends on experimental parameters like the oscillatory mode. Training with stress-controlled cyclic deformations after a conditioning step constitutes the fastest and most efficient annealing conditions with little sensitivity to the training stress. Detailed investigations have been carried out and will be reported in a more detailed contribution.

A last question concerns the generality of our findings. Figure 6 shows results obtained for a colloidal gel with thixotropic properties, similar to the gel investigated in (13). When the sample is prepared using a traditional preshear protocol at constant rate, we observe tremendous directional memory as revealed by stress-growth experiments performed in the forward and backward directions after flow cessation, as shown in the inset of Fig. 6. Actually this material has even more tendency to develop long-lived residual stress than the SPG. Since in this system the yield point is not well-defined and time dependent, annealing is performed using a reverse amplitude sweep in stress-controlled mode. After annealing, the symmetry upon flow reversal is perfectly restored, as shown in the main panel of Fig. 6. This result supports the general framework proposed in this paper, based on the control of the distribution of local stresses, irrespective of the microscopic details of the sample, and opens the way to further investigations aimed at assessing the minimal requirements for the successful applicability of the proposed experimental protocol. We also believe that our annealing protocol should be successful when applied to many different amorphous materials with various microstructures, provided that they have the capacity to store local stresses and strains. This prompts further investigations combining experiments and simulations.

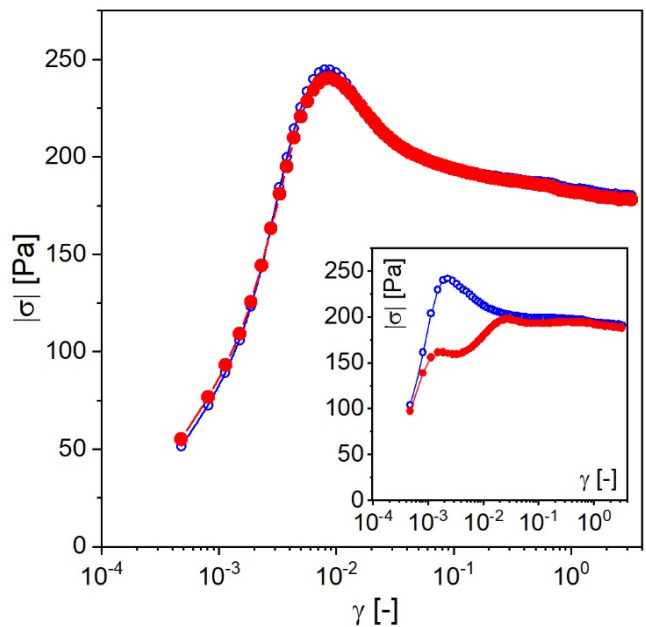

**Fig. 6** Experimental stress-growth experiments in the forward direction (blue) and backward direction (red) for the fume silica gel after optimum annealing. The inset shows the results without annealing.

## Materials and Methods

**Microgels**

The microgels used in this study are synthesized by COATEX SAS. The particles are well-defined spherical networks, made of ethyl acrylate, methacrylic acid and diallyl phthalate as a crosslinker. Suspensions are prepared by diluting the stock solution (30 wt%) with ultrapure water and adding sodium hydroxide that ionizes the methacrylic acid units and provokes the swelling of the particles ($R_h \cong 165$ nm). The suspensions undergo a jamming transition at $C \cong$ 1 wt% above which they become athermal yield stress materials that behave like solids at rest but flow when the stress is large enough. In the jamming regime, the particles are packed in a glassy state and interact through elastic forces. The concentration of the suspensions used in this study is $C \cong 2$ wt%; they have a low-frequency storage modulus $G_0 = 435$ Pa, a yield stress $\sigma_y = 32 \pm 2$ Pa and a yield strain $\gamma_y = 0.075 \pm 0.003$ (*SI Appendix,* Table S1).

**Colloidal gel** It is prepared from fumed silica in an organic solvent. Fumed silica is kindly donated by EVONIK. The gel is prepared in conditions similar to (13). The powder is dispersed at a concentration of 6.7 wt% wt in a mixture of paraffin oil (70 wt%) and polybutylene (30 wt%). The dispersion is homogenized by mixing overnight with an overhead stirrer (IKA RW16) at a rate of 10 s$^{-1}$ at a temperature of 80 °C.

**Rheological measurements**

We use an Anton Paar MCR 502 equipped with the firmware FM4.92, working either in Controlled Rate mode (CR) or Controlled Stress mode (CS). The geometry is a 50 mm diameter cone and plate, with rough surfaces (5 µm) to avoid the occurrence of wall slip. To minimize solvent evaporation, the geometry is protected from the exterior by a solvent trap commercially available from Anton Paar. The atmosphere of the trap is saturated using a small quantity of water. Three main rheological tests are performed: (a) strain sweep experiments consist in subjecting the material to an oscillatory shear deformation at constant frequency ($\omega = 1$ Hz) and stress amplitude increasing from $10^{-3}$ to $10^{2}$. (b) stress-growth experiments or start-up flow experiments consists in applying a constant shear rate $\dot{\gamma}_0$ from rest; (c) flow curves are built by applying shear rates between $10^{3}$ and $10^{-3}$ s$^{-1}$ and recording the stress value when steady state is reached. All experiments are performed at 20 °C.

**Simulation Method**

We perform molecular dynamics (MD) simulations of a model Soft Particle Glass (SPG), made of slightly polydisperse deformable spheres with Hertzian contacts, dispersed in a solvent with a viscosity $\eta_S$. The SPG consists of $N = 10^5$ particles, with an average radius $R$, confined in a cubic box where they are jammed at a volume fraction $\phi = 0.84$. Initially, the glass is equilibrated in a state where the force and torque on each particle are zero. Oscillatory shear flow, start-up flow, and continuous flow are generated using Lees–Edwards boundary conditions in the LAMMPS package (36). Two particles denoted by $\alpha$ and $\beta$ interact through repulsive contact forces of elastic origin ($\mathbf{f}^e_{\alpha\beta}$) and elastohydrodynamic lubrication forces ($\mathbf{f}^{EHD}_{\alpha\beta}$). The equation of motion are made dimensionless by scaling time, stress, and length by $\eta_S/E^*$, $E^*$, and $R$ respectively. The dynamics is thus solely characterized by the dimensionless shear rate $\dot{\gamma}\eta_S/E^*$ (steady flow) and dimensionless frequency $\omega\eta_S/E^*$ (oscillatory flow). $E^* = E/2(1-\nu^2)$ is the contact modulus ($E$: Young modulus; $\nu$: Poisson ratio). A description of the mathematical formulation and a derivation of the forces acting on the particles can be found in previous publications (4, 29, 31). From the interparticle forces and particle positions ($\mathbf{x}_\alpha$ and $\mathbf{x}_\beta$), we implement the Kirkwood formula (37) to determine both the average stress and per-particle stress. The average stress of the suspension is computed as:

$$\bar{\boldsymbol{\sigma}} = \frac{1}{V}\sum_{\beta}^{N}\sum_{\alpha>\beta}^{N}\left(\mathbf{f}^e_{\alpha\beta} + \mathbf{f}^{EHD}_{\alpha\beta}\right)\left(\mathbf{x}_\alpha - \mathbf{x}_\beta\right)$$

where $V$ is the volume of the simulation box. The per-particle stress is calculated by individually accounting for the contribution of each particle to the total stress as follows:

$$\boldsymbol{\sigma}_p = \frac{N}{V}\sum_{\beta}^{N}\left(\mathbf{f}^e_{\alpha\beta} + \mathbf{f}^{EHD}_{\alpha\beta}\right)\left(\mathbf{x}_\alpha - \mathbf{x}_\beta\right)$$

The probability distribution function $P(\sigma_p/E^*)$ of the per-particle shear stress is obtained by creating a histogram by binning the stress values within the range obtained from the calculations. $P(\sigma_p/E^*)$ is analyzed in terms of its first moments $\mu_i$, which give access to the mean value ($\mu_1$), standard deviation ($\mu_2^{1/2}$) and skewness ($S = \mu_3/\mu_2^{3/2}$).